# Energy-conserving molecular dynamics is not energy conserving

Lina Zhang,[†] Yi-Fan Hou,[†] Fuchun Ge, Pavlo O. Dral[*]

*State Key Laboratory of Physical Chemistry of Solid Surfaces, College of Chemistry and Chemical Engineering, Fujian Provincial Key Laboratory of Theoretical and Computational Chemistry, and Innovation Laboratory for Sciences and Technologies of Energy Materials of Fujian Province (IKKEM), Xiamen University, Xiamen, Fujian 361005, China*

E-Mail: dral@xmu.edu.cn

**Abstract**

Molecular dynamics (MD) is a widely-used tool for simulating the molecular and materials properties. It is a common wisdom that molecular dynamics simulations should obey physical laws and, hence, lots of effort is put into ensuring that molecular dynamics simulations are energy 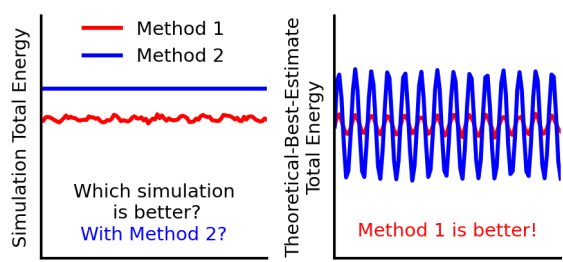 conserving. The emergence of machine learning (ML) potentials for MD leads to a growing realization that monitoring conservation of energy during simulations is of low utility because the dynamics is often unphysically dissociative. Other ML methods for MD are not based on a potential and provide only forces or trajectories which are reasonable but not necessarily energy-conserving. Here we propose to clearly distinguish between the simulation-energy and true-energy conservation and highlight that the simulations should focus on decreasing the degree of true-energy non-conservation. We introduce very simple, new criteria for evaluating the quality of molecular dynamics estimating the degree of true-energy non-conservation and we demonstrate their practical utility on an example of infrared spectra simulations. These criteria are more important and intuitive than simply evaluating the quality of the ML potential energies and forces as is commonly done and can be applied universally, e.g., even for trajectories with unknown or discontinuous potential energy. Such an approach introduces new standards for evaluating MD by focusing on the true-energy conservation and can help in developing more accurate methods for simulating molecular and materials properties.

[†] Co-first authors

Natural phenomena obey the fundamental law of conservation of energy, according to which, the exact total energy of an isolated atomistic system with no exchange of particles and energy with the environment should stay precisely the same over time[1]. For a computer simulation to rigorously obey this law and accurately reproduce natural phenomena, the simulation should ensure that the exact total energy stays the same over time. This would require exactly solving the Schrödinger (in non-relativistic approximation) or Dirac equation (for relativistic treatment) in dynamics simulations.[2, 3] This is only possible for the simplest systems and thus, simulations are usually performed using approximations such as treating nuclei classically and propagating them using Newtonian laws as is typically done in molecular dynamics (MD).[4, 5]

The propagation requires knowledge of forces acting on nuclei which are commonly calculated as the negative of derivatives of potential energy defined either quantum mechanically (QM) or with some other empirical analytic expression (molecular mechanics force field or machine learning (ML) potential). In such MD simulations, the total energy $E_{tot}$ is calculated as the sum of the kinetic energy of classical nuclei $E_{kin}$ and the potential energy of the entire system $E_{pot}$ evaluated by a given atomistic model[6, 7]:

$$E_{tot} = E_{pot} + E_{kin}. \qquad (1)$$

Conventionally, MD is said to be energy conserving if the total energy stays the same during simulations. Nevertheless, this total energy is only an approximation to the unknown exact total energy and the law of conservation of energy refers to the conservation of the exact total energy. Here we highlight the need to clearly distinguish these two concepts. The total energy evaluated by Eq. 1 is recommended to be referred to as simulation total energy and if it is unchanged during MD, dynamics can be said to be simulation-energy conserving. On the other hand, for the dynamics conserving the exact total energy we can use the term 'true-energy conservation'. We cannot use the term 'exact-energy conservation' because it often refers to 'exact conservation' rather than to 'exact energy'.

In this work we show how making a distinction between the simulation-energy and true-energy conservations allows us to better gauge the quality of MD. Particularly, we demonstrate that estimating how much dynamics deviates from the true-energy conservation gives an intuitive and unique insight into the MD quality compared to other widely used approaches, e.g., evaluating the deviations from the simulation-energy conservation or the accuracy of potential energies or forces given by an atomistic model.

This distinction is essential for ML potentials that are known to provide simulation-energy conserving dynamics but which can be nonetheless unphysical, e.g., lead to exploding molecules.[8, 9] It also provides us with a consistent way of estimating atomistic models which do not provide energies, e.g., for ML models which only predict forces but no energies[10] and for novel models which directly predict nuclear coordinates evolution by evaluating neither energies nor forces.[11]

Before we move on to pure ML models, we highlight general considerations applicable to any model, either QM or ML. We start by providing a concrete example where simulation-energy conservation does not by itself ensure that the dynamics is of satisfactory quality. MD is a powerful tool for simulating infrared (IR) spectra. We propagated MD trajectories for the $N_2O$ molecule with three different atomistic models: fast and accurate ML-improved semi-empirical QM method AIQM1,[12] a popular density functional theory method with moderate computational cost PBE[13]/def2-SVP[14], and fast semi-empirical QM method GFN2-xTB[15] (see Methods for details). All three MD trajectories are simulation-energy conserving (Figure 1b) but provide spectra of very different quality (Figure 1a). Compared to the experiment[16], ML-based AIQM1 gives the best agreement in peak positions and a reasonable agreement in intensities, PBE/def2-SVP peaks are shifted to the higher frequencies by ca. 100 cm$^{-1}$ and GFN2-xTB has the largest shift by as much as 300 cm$^{-1}$.

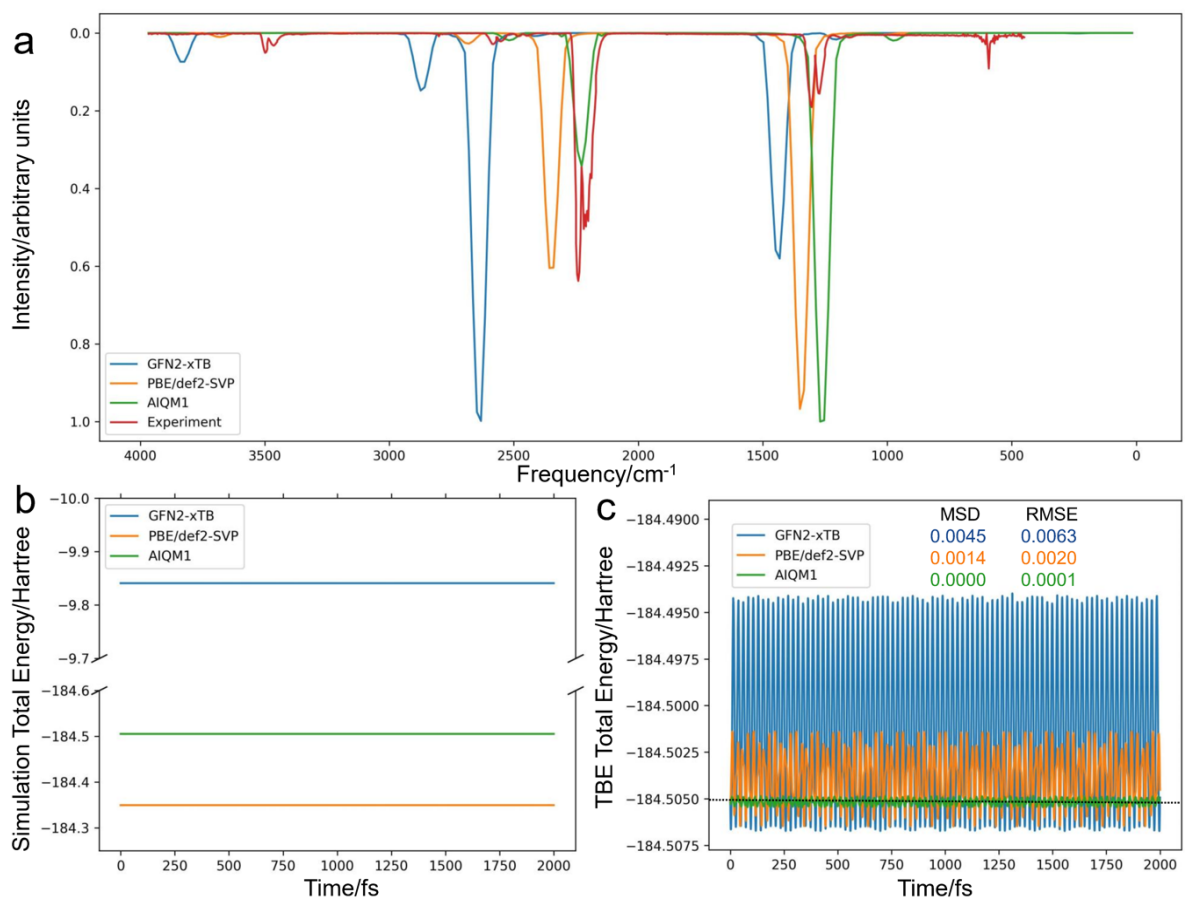

**Figure 1.** Molecular dynamics of nitrous oxide (N$_2$O) started with the same initial conditions and propagated with 0.50 fs time step and different atomistic models: GFN2-xTB (blue), PBE/def2-SVP (orange), and AIQM1 (green). a) Simulated and experimental[16] (red) infrared spectra. Evolutions of b) simulation total energy and c) theoretical-best estimate (TBE) of total energy. MSD (mean signed deviation) and RMSE (root-mean-squared error) are calculated for deviations of TBE total energies during dynamics relative to time zero (shown with a dashed line).

In our example, the different quality of MD is obviously the result of the different accuracy of the used atomistic models. Here we suggest evaluating this accuracy by estimating how much the trajectories deviate from the true-energy conservation. For this, we need to find a computationally efficient approximate theoretical-best estimate (TBE) for the unknown exact total energy. In the context of the classical and Born–Oppenheimer MD, we propose to evaluate the TBE total energy $E_{\text{tot}}^{\text{TBE}}$ along MD by a straightforward modification of Eq. 1 for a given time step:

$$E_{\text{tot}}^{\text{TBE}} = E_{\text{pot}}^{\text{TBE}} + E_{\text{kin}}, \qquad (2)$$

where $E_{\text{kin}}$ is trivially the same in both Eqs. 1 and 2 because it just depends on the velocities and masses of nuclei in a given trajectory. An important difference in Eq. 2 compared to Eq. 1 is that the potential energy $E_{\text{pot}}^{\text{TBE}}$ is obtained by performing single-point calculations with the TBE atomistic model as a post-processing step after dynamics propagation. The TBE atomistic model should be as accurate as possible but it should be still computationally affordable. The TBE total energy can be calculated in parallel (while trajectory can only be propagated sequentially) and only for a subset of time steps to further reduce the cost.

For the $N_2O$ dynamics, we use the gold-standard CCSD(T)*/CBS method[17] as the TBE approach (see Methods for details). Analysis of the time evolution of the TBE total energy provides a simple and intuitive overview of the trajectory quality. The better the quality of MD, the less TBE total energy deviates from the straight line which can be drawn from the TBE total energy calculated for the initial conditions which are the same for all trajectories (Figure 1c). Thus, quick visual inspection helps us to detect issues with MD and in our example, it is obvious that the TBE total energy is much closer to being constant for ML-based AIQM1 and the largest fluctuations are for the GFN2-xTB, following the trend from our previous comparison to experiment (Figure 1). We can thus estimate that the AIQM1 trajectory has the least deviation from the true-energy conservation which leads to a better agreement with experiment.

The qualitative analysis can be readily extended to the quantitative error measures. We can use the non-conservation of TBE total energy to get an estimate of systematic and non-systematic deviation by using the root-mean-squared error (RMSE) and the mean signed deviation (MSD) calculated for the entire trajectory as the deviation of TBE total energy from the initial TBE total energy at time zero:

$$\text{RMSE} = \sqrt{1/\text{number of time steps} \sum \left(E_{\text{tot}}^{\text{TBE}}(\text{time step}) - E_{\text{tot}}^{\text{TBE}}(\text{time zero})\right)^2}, \quad (3)$$

$$\text{MSD} = 1/\text{number of time steps} \sum \left(E_{\text{tot}}^{\text{TBE}}(\text{time step}) - E_{\text{tot}}^{\text{TBE}}(\text{time zero})\right). \quad (4)$$

The RMSE and MSD values for the above three trajectories of $N_2O$ are clearly in line with the quality of the spectra obtained from MD (Figure 1). MSD values also indicate systematically too-high total energy in the simulations with the GFN2-xTB model which corresponds to the too-high frequency observed in the spectrum obtained with this model.

This example highlights that estimating deviations from true-energy conservation with TBE total energies gives a straightforward way to judge the quality of MD simulations with any model (ML or QM), e.g., IR spectra which can be useful in absence of experimental spectra and when a computational cost of using the high-level TBE model to run dynamics is too high. Also, qualitative aspects of dynamics such as too-high frequencies can be nicely explained by systematically too-high TBE total energies. Too high TBE total energy means that the system has systematically more energy than it had at time zero and thus dynamics unphysically gains energy from the vacuum and corresponds to too-hot dynamics.

The goal of improving the quality of MD simulations can be then framed as approaching closer to the true total energy conservation. This allows us to shift the focus from simulation-energy conservation, which does not ensure the quality of the simulation, to more important true-energy conservation, directly connected to the simulation quality. This shift of focus also provides a simple approach to validate and justify possible alternative strategies which may be simulation-energy non-conserving but have smaller deviations from true-energy conservations.

This is particularly important for ML potentials as we demonstrate on another concrete example: MD of the hydrogen molecule for which we propagate with fast pure ML potential ANI-1ccx[17], hybrid QM/ML method AIQM1, slow *ab initio* QM method RMP2[18, 19], and very accurate and slow FCI (full configuration interaction, recovering exact energy for $H_2$ within Born–Oppenheimer approximation and time-independent non-relativistic Schrödinger equation)[20, 21]. We use FCI trajectory as a reference and perform FCI single-point calculations on other trajectories to evaluate TBE total energies. We propagate trajectories with a small time step of 0.05 fs and one additional trajectory with 0.5 fs time step with the RMP2 model.

All trajectories but one (with 0.5 fs time step) are clearly simulation-energy conserving and if we just analyze the simulation total energy, we would arrive at a conclusion that the trajectory with 0.5 fs time step is the worst (Figure 2a). This will be, however, a very wrong conclusion. Analysis of the time-evolution of internuclear distances shows that the simulation-energy non-conserving trajectory with 0.5 fs time step is among the best compared to the reference FCI trajectory (Figure 2c). ML potential ANI-1ccx on the other hand is simulation-energy conserving but its trajectory for $H_2$ is clearly the worst as the molecule dissociates. This problem of unphysically unstable simulations is common for simulation-energy-conserving ML potentials.[8, 9] All of this is mirrored in the analysis of the TBE total energy, which has on average the smallest deviations from the straight line in the case of the MP2 trajectories and the largest – for ML potential ANI-1ccx (Figure 2b). All trajectories have systematically too-high TBE total energies which provide a simple interpretation for the observed too-high frequency of $H_2$ vibrations or dissociation.

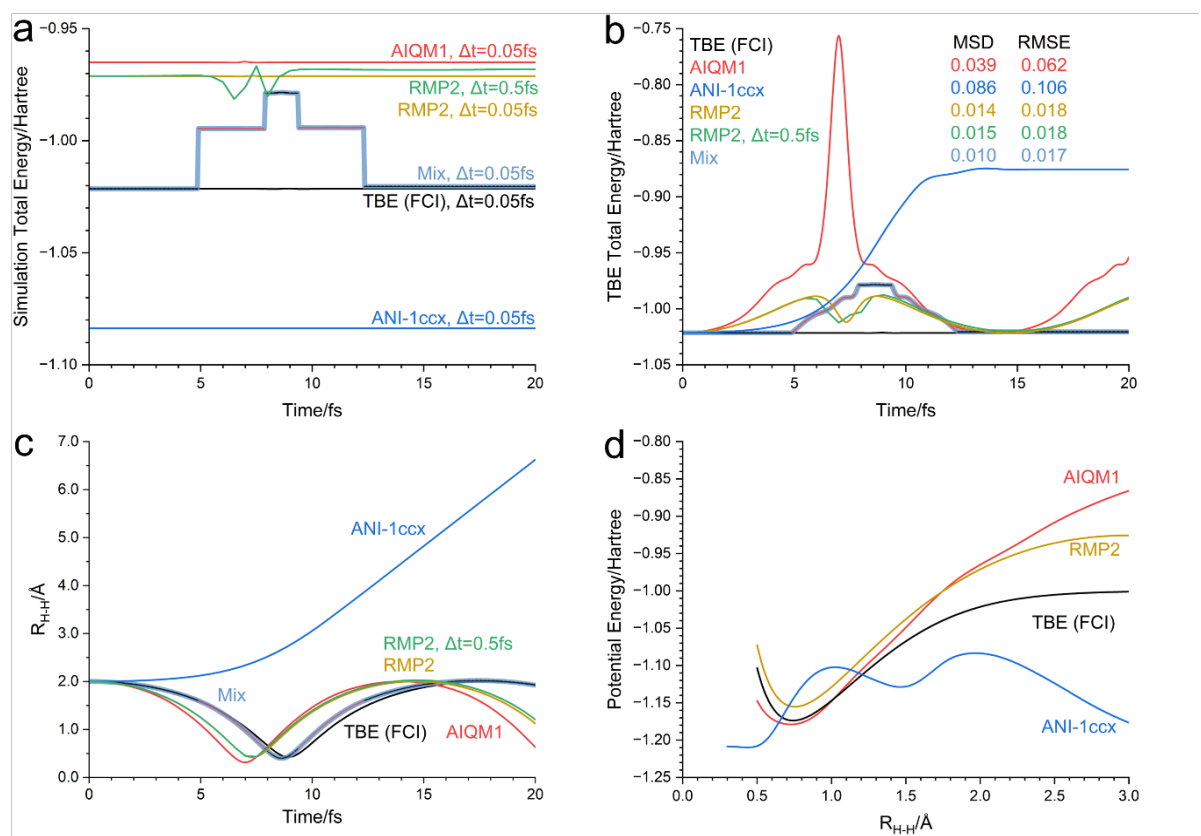

**Figure 2.** Molecular dynamics trajectories of $H_2$ started with the same initial conditions and propagated with 0.05 fs time step and different models: TBE (FCI, black), AIQM1 (red), ANI-1ccx (blue), RMP2 (yellow), Mix (livid, the FCI and AIQM1 parts are marked with colors corresponding to models). An additional trajectory at RMP2 propagated with 0.5 fs time step is also shown (green). a,b) Evolutions of simulation and TBE total energies (lines of colors corresponding to models). c) Evolutions of internuclear distances. d) Potential energies along the internuclear distance. MSD (mean signed deviation) and RMSE (root-mean-squared error) are calculated for deviations of TBE total energies during dynamics relative to time zero (shown with a dashed black line).

We can also recast our goal of approaching closer to the true total energy conservation as a more concrete goal of having as conservative as possible TBE total energy, rather than simulation total energy (which is in many cases also desirable to ensure numerical stability of MD[7]). Unfortunately, we cannot always choose more accurate atomistic models for MD as they often come at a prohibitive cost. Thus, we propose a simple mixed-model strategy that whenever necessary switches to an appropriate model that has satisfactory accuracy. Similar approaches were used in previous studies[22-26] and here we provide justification from the perspective of true (approximated by TBE) energy conservation.

Building upon our previous example of $H_2$, we use computationally cheap ML-based AIQM1, where it works well (for bond lengths between 0.6 and 1.6 Å), and switch to costly FCI for regions, where AIQM1 is not so good (based on PES analysis, Figure 2d). This mixed-model MD obviously breaks the continuity of simulation total energy (Figure 2a). However, the TBE total energy of the mixed-model trajectory has on average smaller deviation from the constant value than the pure AIQM1 trajectory both visually and quantitively (MSD has dropped from 0.039 to 0.010 Hartree, Figure 2b). The bond length in the mixed-model trajectory very closely follows the pure TBE (FCI) trajectory, while the pure AIQM1 trajectory has too fast oscillations (Figure 2c). The cost of this mixed-model strategy is much lower than that of pure high-level MD run with pure FCI, while the accuracy is comparable.

We can also combine two approximate (non-TBE) models to achieve better results in terms of TBE total energy compared to pure approximate models; this is useful when TBE is too expensive to be used in dynamics. We show this on an example of benzene, where we choose such initial conditions that the ML potential ANI-1ccx leads to dissociation and RMP2 calculations are reasonably accurate but much, ca. 73 times, slower (Figure 3c,d; we use RQCISD[27]/cc-pVDZ[28, 29] method to evaluate TBE total energies). In this case, we switch from ANI-1ccx to RMP2 calculations when the critical C4–H10 bond length exceeds 2.0 Å (Figure 3c,d). Such a mixed strategy has also discontinuities in simulation total energy (Figure 3a) but the TBE total energy has much better behavior and smaller deviation from conservation than in the pure ANI-1ccx trajectory (Figure 3b). Interestingly, the mixed-model trajectory has, on average, a smaller MSD (−0.001 Hartree) than both ANI-1ccx (0.109 Hartree) and RMP2 (−0.008 Hartree) trajectories. The cost compared to pure RMP2 trajectory is halved (for 200 fs). Benzene example shows that the mixed-model strategy, which abandons simulation-energy conservation, can prevent catastrophic, unphysical events observed in simulation-energy-conserving trajectory by switching to a better model when needed.

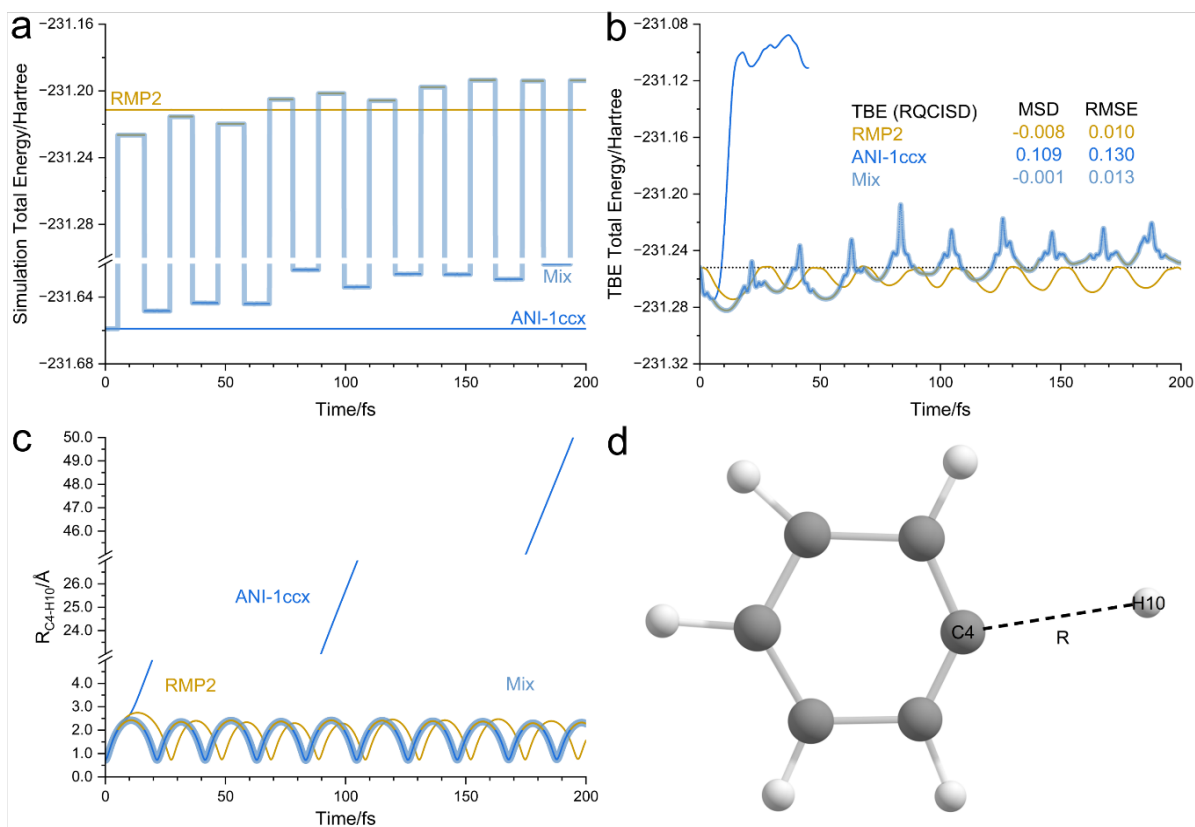

**Figure 3.** Molecular dynamics trajectories of benzene started with the same initial conditions and propagated with 0.05 fs time step and different models: RMP2 (yellow), ANI-1ccx (blue), Mix (livid, the RMP2 and ANI-1ccx parts are marked with colors corresponding to models). a,b) Evolutions of simulation and TBE total energies (lines of colors corresponding to models, black dashed line shows the TBE total energy at time zero). c) Evolution of the C4–H10 bond length. d) The dissociated structure example of benzene during ANI-1ccx dynamics. MSD (mean signed deviation) and RMSE (root-mean-squared error) are calculated for deviations of TBE total energies during dynamics relative to time zero (shown with a dashed black line).

To summarize, here we suggest clearly distinguishing between simulation-energy conservation, commonly considered in dynamics, and true-energy conservation, which was not paid enough attention to but is of great importance in emerging field of ML methods for MD. We argue that the focus of research to ensure simulation-energy conservation is to a large degree insufficient and better improvements can be achieved by targeting true-energy conservation. We propose gauging the true-energy conservation with a concept of theoretical-best-estimate (TBE) total energy conservation which is straightforward to evaluate with the highest-level affordable model for energies. TBE forces are not needed and energy calculations can be performed independently for a small sample of time steps and in parallel. In context of ML potentials, it may be convenient to choose the reference potential used to generate the training data for ML as TBE; this allows to gauge the quality of ML potential in performing real MD simulations rather than by simply benchmarking their accuracy in potential energies and forces as is commonly done[30, 31]. Analysis of the TBE total energies may be used as a standard universal way to evaluate MD trajectories regardless how they were obtained, e.g., directly predicted by ML[11, 32, 33] or by propagating MD with forces evaluated as negative derivatives of ML potential (simulation-energy-conserving MD)[30, 31] or with forces directly predicted by ML (simulation-energy non-conserving MD)[10]. and choose an appropriate affordable model and simulation strategy (Figure 4). We also demonstrated in an example of a mixed-model strategy that we can achieve dynamics, which is not necessarily simulation-energy conserving, but has better true-energy conservation and is physically more correct than simulation-energy-conserving dynamics with some models.

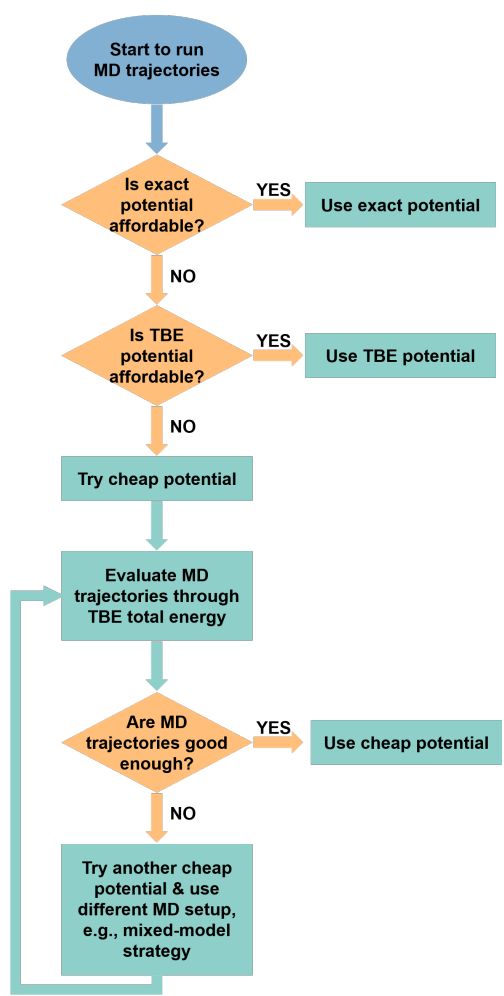

**Figure 4.** Schematic representation of strategies for better dynamics considering true-energy conservation rather than simulation energy conservation.

**Methods**

*Computational details*

AIQM1[12] calculations were performed with the development version of MLatom program[34-36] interfaced to the MNDO program[37] (providing the ODM2*[12, 38] part of AIQM1), TorchANI 2.2[39] (providing modified ANI-type neural-network part of AIQM1), and dftd4[40] (providing the D4-part[41] of AIQM1). CCSD(T)*/CBS is performed with MLatom interfaced to the ORCA 4.2.0 software package[42, 43]. CCSD(T)* is done by approximation to CCSD(T) based on domain-based local-pair natural-orbital-CCSD(T)[44] multi-step calculations and CBS is an extrapolation to complete basis set following the scheme described before[17]. ANI-1ccx[17] calculations were also performed with MLatom interfaced to TorchANI 2.2. and Gaussian 16[45] (used for geometry optimization). PBE (with the def2-SVP basis set), MP2 (with the cc-pVDZ basis set), FCI (with the cc-pVQZ[28, 29] basis set), and RQCISD (with the cc-pVDZ basis set) calculations were done with the Gaussian program[45]. GFN2-xTB calculations were performed with the xtb program.[46] Molecular dynamics and power spectra calculations were performed with MLatom (the theory is described below). Initial conditions for hydrogen were chosen manually (starting at 2.0 Å and zero initial velocities) and for benzene by sampling from Wigner distribution and choosing conditions with the large initial total energy to cause the dissociation event in ANI-1ccx. The Wigner sampling was performed using the Newton-X's initcond routine[47].

*Molecular dynamics*

The molecular dynamics codes used in this research are implemented in MLatom package[34-36], which supports calculations with many ML models such as ANI-1ccx[17], AIQM1[12], sGDML[48, 49], KREG[50, 51], etc., and QM models, often, via interfaces to third-party software (required third-party software for this study are listed in Computational details above). For some models (e.g., ANI-1ccx), the MD can operate data in RAM reducing the need for disk I/O and thus ensuring fast simulation speed.

In the NVE ensemble, our implementation uses the velocity Verlet algorithm[52] which propagates MD trajectory with the following equations:

$$\mathbf{x}(t + \Delta t) = \mathbf{x}(t) + \mathbf{v}(t)\Delta t + \frac{1}{2}\mathbf{a}(t)\Delta t^2$$

$$\mathbf{v}(t + \Delta t) = \mathbf{v}(t) + \frac{1}{2}[\mathbf{a}(t) + \mathbf{a}(t + \Delta t)]\Delta t,$$

where **x** is the coordinate, **v** – the velocity, **a** – the acceleration, $t$ – the time and $\delta t$ – the time step. This algorithm for models which are energy-conservative by construction and small time steps is approaching simulation total energy conservation during the MD simulations in the NVE ensemble.

MLatom also supports simulations in the NVT ensemble (also known as canonical ensemble) with Nosé–Hoover chain, where the velocities of particles are controlled by a chain of additional degrees of freedom to guarantee the canonical sampling of the original system. The equations of motion are shown below[53]:

$$\mathbf{v}_i = \frac{\mathbf{p}_i}{m_i},$$

$$\dot{\mathbf{p}}_i = \mathbf{F}_i - \mathbf{p}_i \frac{p_{\xi_1}}{Q_1},$$

$$\xi_i = \frac{p_{\xi_i}}{Q_i},$$

$$\dot{p}_{\xi_1} = \left(\sum_i \frac{\mathbf{p}_i^2}{m_i} - N_f kT\right) - p_{\xi_1} \frac{p_{\xi_2}}{Q_2},$$

$$\dot{p}_{\xi_k} = \left(\frac{p_{\xi_{k-1}}^2}{Q_{k-1}} - kT\right) - p_{\xi_k} \frac{p_{\xi_{k+1}}}{Q_{k+1}},$$

$$\dot{p}_{\xi_M} = \left(\frac{p_{\xi_{M-1}}^2}{Q_{M-1}} - kT\right).$$

The $\mathbf{v}_i$, $\mathbf{p}_i$ and $m_i$ are the velocity, momentum and mass of atom $i$ in the system and $\mathbf{F}_i$ is the force experienced by this atom. The chain of additional degrees of freedom with length $M$ is applied to the system, which have the coordinate $\xi_k$, the momentum $p_{\xi_k}$ and the mass $Q_k$. $N_f$ is the degrees of freedom of the system, $k$ the Boltzmann constant, and $T$ the temperature. The choice of $Q_k$ can be chosen by the user but should follow some general rules[54, 55]:

$$Q_1 = N_f kT/\omega_p^2,$$

$$Q_k = kT/\omega_p^2,$$

where $\omega_p$ is the fluctuation frequency of the thermostat and it should be comparable to vibrational frequencies or phonon frequencies one wish to equilibrate.

In our implementations, we followed the work by Martyna et al.[56], where the Liouville approach is used to integrate the equations of motion. The Liouville operator of this extended system ($L$) is first separated into operators of positions ($L_x$), velocities ($L_v$), and the Nosé–Hoover thermostats ($L_{\text{NHC}}$):

$$iL = iL_x + iL_v + iL_{\text{NHC}}.$$

Then, we factorize $iL$ using the Trotter formula to get:

$$e^{iL\Delta t} = e^{iL_{\text{NHC}}\Delta t/2} e^{iL_v\Delta t/2} e^{iL_x\Delta t} e^{iL_v\Delta t/2} e^{iL_{\text{NHC}}\Delta t/2} + \mathcal{O}(\Delta t^3).$$

To obtain as accurate a trajectory as we can, we use a higher-order integrator applied to the evolution of operator $L_{\text{NHC}}$[56-58]:

$$e^{iL_{\text{NHC}}\Delta t/2} = \prod_{i=1}^{n_c}\left[\prod_{j=1}^{n_{ys}} e^{iL_{\text{NHC}} w_j \Delta t/2n_c}\right] + \mathcal{O}(\Delta t/2n_c)^5,$$

where $n_c$ is the multiple time step, $n_{ys}$ is the number of the Yoshida–Suzuki steps, and $w_j$ are the Yoshida–Suzuki coefficients[57, 58]. The expression for the approximation of $e^{iL_{\text{NHC}} w_j \Delta t/2n_c}$ is taken from the Ref. [56]:

$$\begin{aligned}
&e^{iL_{\text{NHC}} w_j \Delta t/2n_c} \\
&= \exp\left(\frac{\Delta t}{4n_c} G_M \frac{\partial}{\partial v_{\xi_M}}\right) \\
&\times \exp\left(-\frac{\Delta t}{8n_c} v_{\xi_M} v_{\xi_{M-1}} \frac{\partial}{\partial v_{\xi_{M-1}}}\right) \exp\left(\frac{\Delta t}{4n_c} G_{M-1} \frac{\partial}{\partial v_{\xi_{M-1}}}\right) \exp\left(-\frac{\Delta t}{8n_c} v_{\xi_M} v_{\xi_{M-1}} \frac{\partial}{\partial v_{\xi_{M-1}}}\right) \\
&\times \cdots \times \exp\left(-\frac{\Delta t}{8n_c} v_{\xi_2} v_{\xi_1} \frac{\partial}{\partial v_{\xi_1}}\right) \exp\left(\frac{\Delta t}{4n_c} G_1 \frac{\partial}{\partial v_{\xi_1}}\right) \exp\left(-\frac{\Delta t}{8n_c} v_{\xi_2} v_{\xi_1} \frac{\partial}{\partial v_{\xi_1}}\right) \\
&\times \exp\left(-\frac{\Delta t}{2n_c} \sum_{i=1}^{N} v_{\xi_1} \mathbf{v}_i \nabla_{\mathbf{v}_i}\right) \exp\left(\frac{\Delta t}{2n_c} \sum_{i=1}^{M} v_{\xi_i} \frac{\partial}{\partial \xi_i}\right) \times \cdots \\
&\times \exp\left(-\frac{\Delta t}{8n_c} v_{\xi_M} v_{\xi_{M-1}} \frac{\partial}{\partial v_{\xi_{M-1}}}\right) \exp\left(\frac{\Delta t}{4n_c} G_{M-1} \frac{\partial}{\partial v_{\xi_{M-1}}}\right) \exp\left(-\frac{\Delta t}{8n_c} v_{\xi_M} v_{\xi_{M-1}} \frac{\partial}{\partial v_{\xi_{M-1}}}\right) \\
&\times \exp\left(\frac{\Delta t}{4n_c} G_M \frac{\partial}{\partial v_{\xi_M}}\right),
\end{aligned}$$

where

$$G_1 = \frac{1}{Q_1}\left(\sum_{i=1}^{N} m_i \mathbf{v}_i^2 - N_f kT\right),$$

$$G_i = \frac{1}{Q_i}\left(Q_{i-1} v_{\xi_{i-1}}^2 - kT\right), for\ i > 1.$$

In MLatom, we support an arbitrary choice of $M$ and $n_c$ and $n_{ys}$ of 1, 3, 5, and 7.

Initial conditions (initial geometries and velocities) can be provided by the user or using one of the built-in additional ways to generate initial conditions for MD simulations. One way is, for a given initial geometry, to generate random velocities with eliminated linear and angular momenta. Users can set the initial kinetic energy or the "initial temperature" $T_{init}$ to control the initial kinetic energy ($N_f kT/2$) of the system with $N_f$ degrees of freedom. The other way is to sample both geometries and velocities from the Wigner distribution (normal mode sampling) with the Newton-X's initcond routine[47].

*Infrared spectrum*

The infrared spectra are calculated using the fast Fourier transform as[59, 60]:

$$P(\omega) \propto \frac{\nu \tanh\left(\frac{h\nu}{2k_B T}\right)}{n(\nu)} \int \langle \boldsymbol{\mu}(\tau)\boldsymbol{\mu}(t+\tau)\rangle_\tau e^{-i2\pi\nu t} dt,$$

where $\langle \boldsymbol{\mu}(\tau)\boldsymbol{\mu}(t+\tau)\rangle_\tau$ is the autocorrelation function of dipole moment. In our implementations, before performing the fast Fourier transform, a Hann window function is applied to the autocorrelation function. The users can tune autocorrelation depth to control the width of peaks. Zero padding can also be applied to the autocorrelation function to improve the resolution.

In the $N_2O$ dynamics, the system was first equilibrated using NVT ensemble (with Nose–Hoover chain mentioned above) at 300 K for 10 ps, where AIQM1 was used for dynamics propagation. We used the final geometry and nuclear velocities from this trajectory as initial conditions to start 23 ps-long trajectories with the NVE ensemble using GFN2-xTB, PBE/def2-SVP, and AIQM1 atomistic models. The IR spectra were generated from the last 20 ps of these trajectories as the first 3 ps were removed.

**Code availability**

All code used in this study was implemented in the development versions of the open-source MLatom program[34-36] and the version able to perform the reported simulations is publicly available at https://github.com/dralgroup/mlatom/tree/energyconservingmd. Some of the ML dynamics (e.g., with ANI-1ccx, GFN2-xTB, and PBE/def2-SVP) can be also reproduced by running them on the MLatom@XACS cloud computing service currently free for academic and non-commercial uses (https://XACScloud.com).

**Data availability**

The data that support the findings of this study are openly available in figshare at https://doi.org/10.6084/m9.figshare.22315147.

**Acknowledgments**

P.O.D. acknowledges funding by the National Natural Science Foundation of China (No. 22003051 and funding via the Outstanding Youth Scholars (Overseas, 2021) project), the Fundamental Research Funds for the Central Universities (No. 20720210092), and via the Lab project of the State Key Laboratory of Physical Chemistry of Solid Surfaces. This project is supported by Science and Technology Projects of Innovation Laboratory for Sciences and Technologies of Energy Materials of Fujian Province (IKKEM) (No: RD2022070103). The authors also thank Mario Barbatti for his criticism of the early version of the manuscript.

**Author contributions**

Y.F.H. implemented code for molecular dynamics and infrared spectra simulations, and performed calculations of $N_2O$. L.Z. chose the remaining case studies, implemented and carried out all the corresponding simulations. Y.F.H. and L.Z. visualized the results. F.G. did test calculations and debugging as well as contributed to the code for the generation of the initial conditions. P.O.D. conceptualized and wrote the original draft, and took lead in simulation planning. All authors discussed, reviewed, and edited the manuscript and analyzed the results.